\def\ket#1{\left|\, #1\right\rangle}
\def\bra#1{\left\langle #1\right|}
\def\bracket#1#2{\langle #1|#2\rangle}
\def\melement#1#2#3{\langle#1|#2|#3\rangle}
\def\cotimes{\otimes\cdots\otimes}
\def\cH{{\cal H}}
\def\cK{{\cal K}}

\documentclass[12pt]{article}
\usepackage{graphicx,hyperref}
\title{Is there a measurement-only version of quantum mechanics?}
\author{George Svetlichny\footnote{Departamento de Matem\'atica, Pontif\'{\i}cia Universidade Cat\'olica, Rio de Janeiro, Brazil \newline
svetlich@mat.puc-rio.br \hfill
\url{http://www.mat.puc-rio.br/\~svetlich}}}

\begin{document}
\maketitle

\begin{abstract}
Tensor universality often implies that  multi-partite quantum-state processing is determined by what happens in totally disentangled cases. In independent systems relative time direction for the parts is arbitrary. This hints that time may be linked to entanglement and measurements and that there may be a measurement-only version of quantum mechanics. One-way quantum computation suggests that this may be possible.

\end{abstract}

\section{Introduction}

We have recently proved  a theorem \cite{svet:quant-ph/0601093} showing that quantum state processing in entangled systems is uniquely determined by processing in totally disentangles systems. In totally disentangles systems  time direction is arbitrary in the independent parts so one can ask: does time arise from entanglement and measurements? In particular, can unitary time evolution be reduced to entanglement and measurement? This is the reverse side of the quantum mechanical measurement problem. Instead of ``How do measurements happen?" we ask ``How does unitary time evolution happen?". In other words: is there a measurement-only model of quantum mechanics? We argue that ``yes" is a possible answer based on the theory of one-way quantum computation \cite{brow-etal}.

\section{Quantum-state processing and teleportation}
We begin by describing how the existence of teleportation can be
deduced from the behavior of totally disentangled systems.

Consider subjecting a multipartite Heisenberg quantum state
\(\Phi\in \cH_1\cotimes\cH_n\)  to \(m\) successive observations,
described by self-adjoint opertors \(A_1,\dots,A_m\).
 Each \(A_j\) acts in a subproduct of  \(\cH=
\cH_1\cotimes\cH_n\) where we assume it is non-degenerate. Assume
for simplicity that each $A_i$ is a bipartite operator.
Pictorially: (time runs upwards)
\begin{center}
\begin{picture}(110,100)(0,0)
\put(0,0){\line(0,1){58}}\put(0,72){\line(0,1){28}}
\put(25,0){\line(0,1){34}}\put(25,48){\line(0,1){10}}\put(25,72){\line(0,1){28}}
\put(50,0){\line(0,1){34}}\put(50,48){\line(0,1){24}}\put(50,86){\line(0,1){14}}
\put(75,0){\line(0,1){10}}\put(75,24){\line(0,1){48}}\put(75,86){\line(0,1){14}}
\put(100,0){\line(0,1){10}}\put(100,24){\line(0,1){76}}
\put(72,10){\framebox(31,14){ \(A_1\)}}
\put(22,34){\framebox(31,14){ \(A_2\)}}
\put(-3,58){\framebox(31,14){ \(A_3\)}}
\put(47,72){\framebox(31,14){ \(A_4\)}}
\end{picture}
\end{center}

In a given run of the experiment, the state transforms to
\(\Psi=P_m\cdots P_1\Phi\) where each \(P_j\) is a spectral
projection of \(A_j\) in a bipartite subproduct of \(\cH\).

Given a tensor product Hilbert space \(\cH\otimes\cK\) and
\(\Omega\in\cH\otimes\cK\)  one can uniquely define two linear maps
\(g_\Omega:\cH\to \cK^*\) and \(f_\Omega:\cH^*\to \cK\) which in
case  \(\Omega=\alpha\otimes \beta\) act as: $g_\Omega: \ket \phi
\mapsto \bracket\alpha\phi \bra\beta $ and $f_\Omega: \bra \phi
\mapsto \bracket\phi\alpha \ket\beta$.

We illustrate with a pictorial example (time runs upwards):

\begin{center}
\begin{picture}(110,100)(0,0)
\put(0,0){\line(0,1){58}}\put(0,72){\line(0,1){28}}
\put(25,0){\line(0,1){34}}\put(25,48){\line(0,1){10}}\put(25,72){\line(0,1){28}}
\put(50,0){\line(0,1){34}}\put(50,48){\line(0,1){24}}\put(50,86){\line(0,1){14}}
\put(75,0){\line(0,1){10}}\put(75,24){\line(0,1){48}}\put(75,86){\line(0,1){14}}
\put(100,0){\line(0,1){10}}\put(100,24){\line(0,1){76}}
\put(72,10){\framebox(31,14){ \(P_1\)}}
\put(22,34){\framebox(31,14){ \(P_2\)}}
\put(-3,58){\framebox(31,14){ \(P_3\)}}
\put(47,72){\framebox(31,14){ \(P_4\)}}
\end{picture}
\end{center}

Consider a particular example in
\(\cH_1\otimes\cH_2\otimes\cH_3\otimes\cH_4\otimes\cH_5\) given by
\(\Psi=P_4^{(34)}P_3^{(12)}P_2^{(23)}P_1^{(45)}\Phi\) where the
superscripts indicate on which bipartite subproduct the projections
act. Let \(P_i=\ket{\Omega_i}\bra{\Omega_i}\).
 Coeke's theorem \cite{coecke}, here in a slightly modified form, now states that if the initial state is
\(\phi^{\hbox{in}}_1\otimes
\Phi^{\hbox{in}}_{2345}\in\cH_1\otimes(\cH_2\otimes\cH_3\otimes\cH_4\otimes\cH_5),\)
 then the final state is
\(\Phi^{\hbox{out}}_{1234}\otimes\phi^{\hbox{out}}_5\in
 (\cH_1\otimes\cH_2\otimes\cH_3\otimes\cH_4)\otimes\cH_5,\)
where
\begin{equation}\label{coecke}
\phi^{\hbox{out}}_5=g_{\Omega_1}\circ f_{\Omega_4}\circ
g_{\Omega_2}\circ f_{\Omega_3}(\phi^{\hbox{in}}_1).
\end{equation}
Pictorially we have
\begin{center}
\begin{picture}(110,100)(0,0)
\put(0,0){\line(0,1){58}}\put(0,72){\line(0,1){28}}
\put(25,0){\line(0,1){34}}\put(25,48){\line(0,1){10}}\put(25,72){\line(0,1){28}}
\put(50,0){\line(0,1){34}}\put(50,48){\line(0,1){24}}\put(50,86){\line(0,1){14}}
\put(75,0){\line(0,1){10}}\put(75,24){\line(0,1){48}}\put(75,86){\line(0,1){14}}
\put(100,0){\line(0,1){10}}\put(100,24){\line(0,1){76}}
\put(72,10){\framebox(31,14){}} \put(22,34){\framebox(31,14){}}
\put(-3,58){\framebox(31,14){}} \put(47,72){\framebox(31,14){}}
\put(0,29){\vector(0,1){0}} \put(15,58){\vector(1,0){0}}
\put(25,50){\vector(0,-1){0}} \put(40,48){\vector(1,0){0}}
\put(50,60){\vector(0,1){0}} \put(75,48){\vector(0,-1){0}}
\put(65,72){\vector(1,0){0}} \put(90,24){\vector(1,0){0}}
\put(100,62){\vector(0,1){0}} \put(72,10){\framebox(31,14){
\(g_{\Omega_1}\)}} \put(22,34){\framebox(31,14){ \(g_{\Omega_2}\)}}
\put(-3,58){\framebox(31,14){ \(f_{\Omega_3}\)}}
\put(47,72){\framebox(31,14){ \(f_{\Omega_4}\)}}
\end{picture}
\end{center}
where we have put in arrows to indicate the apparent direction of
``quantum information flow". We see that the processing order, given
by the composition of maps in (\ref{coecke}), is not the temporal
order and ``quantum information" switches between ``forward" and
``backward time flow". We consider all the  expressions in quotes in
this paragraph as metaphorical.

An appropriate form of Coecke's theorem holds in any compact closed
category \cite{kell-lapl:JPAA19.193}. Finite dimensional Hilbert spaces are examples of such
categories, so is the category of sets with relations as
morphisms. With sets and maps (which are particular kind of relations),
processing order contrary to temporal
order is not paradoxical. Consider the figure:

\begin{center}
\begin{picture}(220,96)
\put(0,0){\line(0,1){58}}\put(0,70){\line(0,1){26}}
\put(25,0){\line(0,1){26}}\put(25,38){\line(0,1){20}}
\put(25,70){\line(0,1){26}}
\put(50,0){\line(0,1){26}}\put(50,38){\line(0,1){58}}
\put(22,26){\framebox(31,12){\(g\)}}
\put(-3,58){\framebox(31,12){\(f\)}}
\put(65,15){\makebox(0,0){\(\leftarrow\)}}
\put(75,12){\(S\subset X_1\times X_2\times X_3\)}
\put(65,50){\makebox(0,0){\(\leftarrow\)}}
\put(75,47){\(S\cap (X_1\times \Gamma g)\)}
\put(65,85){\makebox(0,0){\(\leftarrow\)}}
\put(75,82){\(S\cap (X_1\times \Gamma g)\cap (\Gamma f\times X_3)\)}
\end{picture}
\end{center}

The vertical lines represent sets and parallel lines are cartesian
products. Each box is a ``filter"  passing  through only those
triples in which the corresponding pair lies on the graph of the
function.

Coecke's theorem is now:  if \(S=\{x\}\times Y\subset X_1\times
(X_2\times X_3)\) then the output, if not empty, is \(\{x\}\times
\{f(x)\}\times \{g(f(x))\}\). Again ``processing order" in
\(g(f(x)\) is not temporal order since the \(g\)-filter acts
first. However, here the boxes commute. ``Filtering" is precisely
like passing particles through various sieves, in the end only those
survive that pass through all of them, regardless of the order. In
the Hilbert space case the boxes do not commute and processing order
is fixed by the topology of the diagram. Its nature contrary to
temporal order seems somewhat mysterious.

It is quite remarkable
that the Hilbert space Coecke's theorem can be proved by considering
only the case in which all relevant states are disentangled, which
becomes similar to the set-theoretic case. To do this we must modify
the state transformer by replacing each projector \(P_\Omega\) by a
general rank-one operator \(
Q_{\Lambda,\Omega}=\ket\Lambda\bra\Omega= \Lambda\otimes
\Omega\rfloor\cdot\) where the floor symbol ``\(\rfloor\)" denotes
the partial inner product defined by \( \Omega\rfloor \alpha\otimes
\beta=(\Omega,\alpha)\beta \).

Consider now the transformation: \[\Psi=Q_m\cdots Q_1\Phi=
\Lambda_m\otimes \Omega_m\rfloor\otimes\cdots \Lambda_1\otimes
\Omega_1\rfloor \Phi.\]
 This is {\em  anti-linear\/} in each \(\Omega_j\) and {\em linear\/} in each \(\Lambda_j\) and in each factor
 of \(\Phi\)  if disentangled from the  rest.

 An example take
\(\Psi=\Lambda_3\otimes \Omega_3\rfloor\, \Lambda_2\otimes
\Omega_2\rfloor\,\Lambda_1\otimes \Omega_1\rfloor\Phi\) defined
pictorially by:
\begin{center}
\begin{picture}(53,144)
\put(0,0){\line(0,1){58}}\put(0,86){\line(0,1){58}}
\put(25,0){\line(0,1){10}}\put(25,38){\line(0,1){20}}
\put(25,86){\line(0,1){20}}\put(25,134){\line(0,1){10}}
\put(50,0){\line(0,1){10}}\put(50,38){\line(0,1){68}}\put(50,134){\line(0,1){10}}
\put(22,10){\framebox(31,14){\(\Omega_1\)}}\put(22,24){\framebox(31,14){\(\Lambda_1\)}}
\put(-3,58){\framebox(31,14){\(\Omega_2\)}}\put(-3,72){\framebox(31,14){\(\Lambda_2\)}}
\put(22,106){\framebox(31,14){\(\Omega_3\)}}\put(22,120){\framebox(31,14){\(\Lambda_3\)}}
\put(0,29){\vector(0,1){0}}
\put(15,58){\vector(1,0){0}}
\put(25,48){\vector(0,-1){0}}
\put(40,38){\vector(1,0){0}}
\put(50,77){\vector(0,1){0}}
\put(36,106){\vector(-1,0){0}}
\put(25,96){\vector(0,-1){0}}
\put(12,86){\vector(-1,0){0}}
\put(0,115){\vector(0,1){0}}
\end{picture}
\end{center}

The arrows will identify processing order, metaphorical ``quantum
information flow".
We now {\sl disentangle everything\/} that is,
\(\Phi=\phi_1\otimes\phi_2\otimes\phi_3\),
\(\Lambda_j=\mu_j\otimes\nu_j\), and
\(\Omega_j=\sigma_j\otimes\tau_j\).

Pictorially:
\begin{center}
\begin{picture}(53,144)
\put(0,0){\line(0,1){58}}\put(0,86){\line(0,1){58}}
\put(25,0){\line(0,1){10}}\put(25,38){\line(0,1){20}}
\put(25,86){\line(0,1){20}}\put(25,134){\line(0,1){10}}
\put(50,0){\line(0,1){10}}\put(50,38){\line(0,1){68}}\put(50,134){\line(0,1){10}}
\put(18,10){\framebox(14,14){\(\sigma_1\)}}\put(43,10){\framebox(14,14){\(\tau_1\)}}
\put(18,24){\framebox(14,14){\(\mu_1\)}}\put(43,24){\framebox(14,14){\(\nu_1\)}}
\put(-7,58){\framebox(14,14){\(\sigma_2\)}}\put(18,58){\framebox(14,14){\(\tau_2\)}}
\put(-7,72){\framebox(14,14){\(\mu_2\)}}\put(18,72){\framebox(14,14){\(\nu_2\)}}
\put(18,106){\framebox(14,14){\(\sigma_3\)}}\put(43,106){\framebox(14,14){\(\tau_3\)}}
\put(18,120){\framebox(14,14){\(\mu_3\)}}\put(43,120){\framebox(14,14){\(\nu_3\)}}
\end{picture}
\end{center}
This depicts three completely independent quantum processes (say, on
Mars, Earth, and Venus).

 The outcome state is:
\[\{ (\sigma_2,\phi_1)\mu_2\} \otimes \{
(\sigma_1,\phi_2)(\tau_2,\mu_1)(\sigma_3,\nu_2)\mu_3\}\otimes \{
(\tau_1,\phi_3)(\tau_3,\nu_1)\nu_3\}.\]

We {\em  rewrite\/} the outcome state by moving the scalar inner
product around:
\[\{(\sigma_2,\phi_1)(\tau_2,\mu_1)(\tau_3,\nu_1)(\sigma_3,\nu_2)\mu_2\}\otimes
\{\mu_3\}\otimes \{\nu_3(\sigma_1,\phi_2)(\tau_1,\phi_3)\}.\]

 This
can be written as (``op" means reverse order in \(\otimes\))
\begin{equation}\label{univ}g_{\Lambda_2}^{\hbox{op}}\circ f_{\Omega_3}^{\hbox{op}} \circ
g_{\Lambda_1}\circ f_{\Omega_2}(\phi_1)\otimes (\Lambda_3\otimes
\Omega_1\rfloor \Phi_{23}). \end{equation}

 This is also {\em  anti-linear\/} in each \(\Omega_j\) and {\em linear\/} in each \(\Lambda_j\) and in each factor
 of \(\Phi\)  if disentangled from the  rest.

 Tensor universality says that two linear (or anti-linear) maps on
\(\cH\otimes\cK\) that coincide on product states are identical.
Thus (\ref{univ})  is  the output state in all cases of \(\Phi\)
being of the form \(\phi_1\otimes \Phi_{23}\) and the \(\Omega_i\)
and \(\Lambda_i\) completely arbitrary. Thus (this case of) Coecke's
theorem is proved.

Let now \(\cH_1\otimes\cH_2\otimes\cH_3\) be a three-qubit Hilbert
space. Alice has access to \(\cH_1\otimes\cH_2\) and Bob to
\(\cH_3\). Let \(\Theta=(\ket0\ket1-\ket1\ket0)/\sqrt 2\), then
\(g_\Theta^{12}\circ f_\Theta^{23}\phi_1=-\frac12\phi_1\) and so
this is an instance of teleportation if by \(\Theta^{23}\) we mean
an entangled pair shared by Alice and Bob (produced, say, by a third
party and invariable), and \(\Theta^{12}\) is an element of a basis
of a two-qubit measurement by Alice. Thus the following diagram
\begin{center}
\begin{picture}(53,96)
\put(0,0){\line(0,1){58}}\put(0,86){\line(0,1){10}}
\put(25,0){\line(0,1){10}}\put(25,38){\line(0,1){20}}
\put(25,86){\line(0,1){10}}
\put(50,0){\line(0,1){10}}\put(50,38){\line(0,1){58}}
\put(22,10){\framebox(31,12){}}\put(22,26){\framebox(31,12){\(\Theta\)}}
\put(-3,58){\framebox(31,12){\(\Theta\)}}\put(-3,74){\framebox(31,12){}}
\end{picture}
\end{center}
teleports.
For other instances of Alice's projection on a measurement
eigenbasis element, she sends classical information to Bob who then
adjusts his qubit by a unitary transformation to achieve
teleportation in all cases. {\em But\/} this teleportation behavior
is a logical consequence of the behavior when all the relevant
states are completely disentangled.
\setlength{\unitlength}{1.5pt}
\begin{center}
\begin{tabular}{ccc}
{\begin{picture}(53,96)
\put(0,0){\line(0,1){58}}\put(0,86){\line(0,1){10}}
\put(25,0){\line(0,1){10}}\put(25,38){\line(0,1){20}}
\put(25,86){\line(0,1){10}}
\put(50,0){\line(0,1){10}}\put(50,38){\line(0,1){58}}
\put(18,10){\framebox(14,12){}}\put(43,10){\framebox(14,12){}}
\put(18,26){\framebox(14,12){\(\mu\)}}\put(43,26){\framebox(14,12){\(\nu\)}}
\put(-7,58){\framebox(14,12){\(\sigma\)}}\put(18,58){\framebox(14,12){\(\tau\)}}
\put(-7,74){\framebox(14,12){}}\put(18,74){\framebox(14,12){}}
\end{picture}} &\raisebox{64pt}{$\quad\Rightarrow\quad$} &
{\begin{picture}(53,96)
\put(0,0){\line(0,1){58}}\put(0,86){\line(0,1){10}}
\put(25,0){\line(0,1){10}}\put(25,38){\line(0,1){20}}
\put(25,86){\line(0,1){10}}
\put(50,0){\line(0,1){10}}\put(50,38){\line(0,1){58}}
\put(22,10){\framebox(31,12){}}\put(22,26){\framebox(31,12){\(\Omega\)}}
\put(-3,58){\framebox(31,12){\(\Lambda\)}}\put(-3,74){\framebox(31,12){}}
\end{picture}}
\end{tabular}
\end{center}
\setlength{\unitlength}{2pt}
Thus from independent quantum processes on Mars, Earth and Venus, we
can deduce that one can teleport a qubit from Rio de Janeiro to Kiev
if one has good enough optical cable and a source of entangled
photons on the Canary Islands.

In totally disentangles systems  time  direction is arbitrary in the
independent parts. Teleportation clearly is not a time reversible
process (there's classical communication involved). So how can
behavior of disentangled system, with weak global temporal
structure, imply that of systems with stronger temporal structure.
Once again one raises the question as to where time comes from, in
particular, does it arise from entanglement and measurements?
\section{One-way quantum computation}

One-way quantum computation \cite{brow-etal} provides one
superficial and apparently affirmative answer to the question posed
at the end of the previous section.

In this approach quantum gates  are simulated by measurement and classical communication, modulo an initial entanglement. Since one can approximate any unitary by a quantum gate circuit, in principle one can simulate unitary time evolution through measurements and classical communication. This would bring us closer to the view that the world is made exclusively of ``events" (measurements). Opposite is the Everett view where there are no events, and  in between is the dualistic Copenhagen view with events (measurement outocmes)and and also event-less unitary evolution. One can thus ask: is the universe a one-way quantum computer? To get some idea of this question, consider the following picture as a typical one-way quantum computation:

\begin{center}
\includegraphics[scale=0.5]{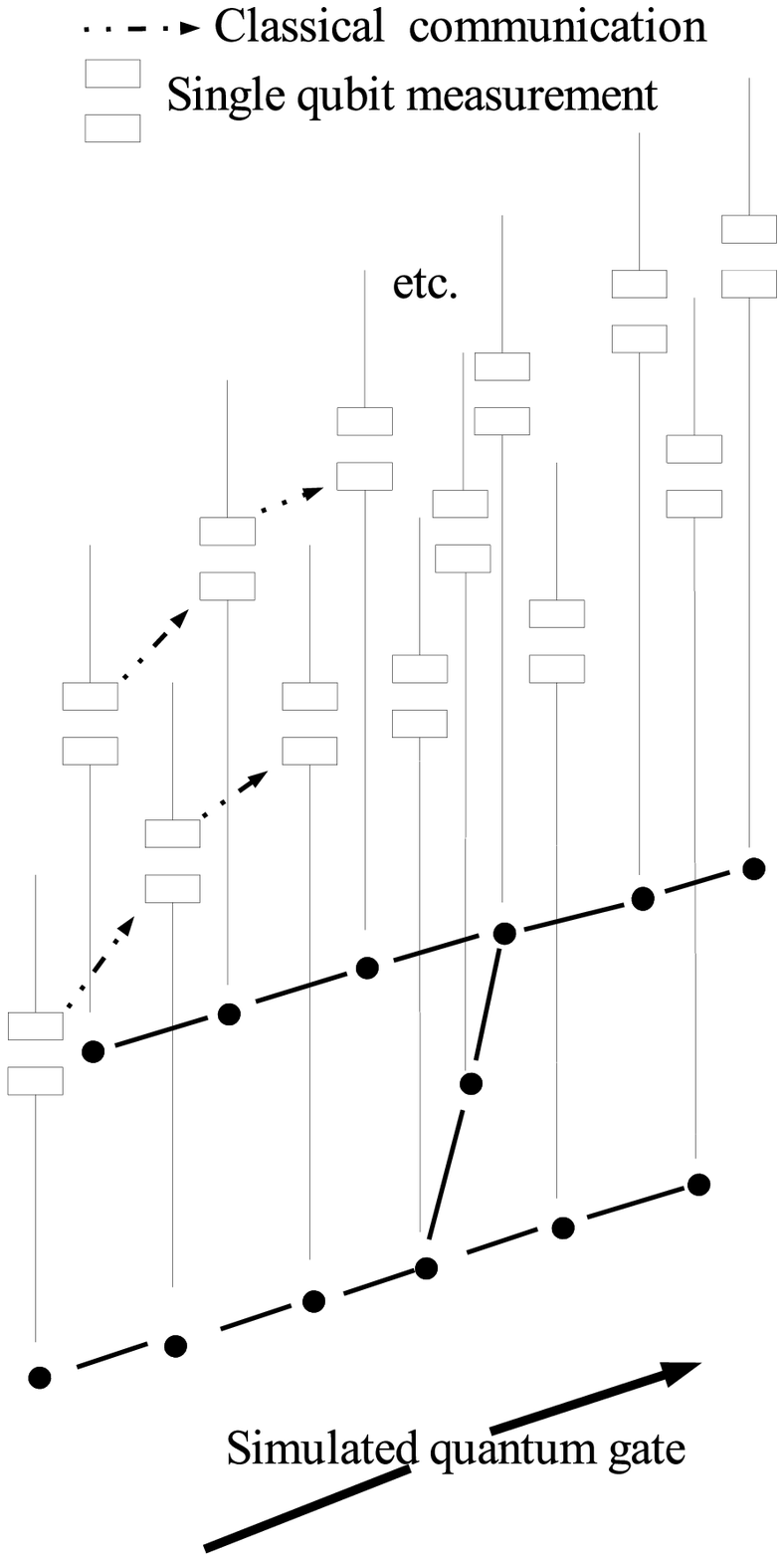}
\end{center}

Here real time runs upward just as in the previous diagrams. Upward
lines are qubit Hilbert spaces. Simulated time runs horizontally.
Thick lines denote entanglement, and what is simulated is a quantum
gate, say CNOT.

To illustrate the essence of one-way computation let us consider how
one would simulate the unitary map \(U\) on a vector
\(\psi=\alpha\ket0+\beta\ket1\), given by
\(U:\alpha\ket0+\beta\ket1\mapsto\alpha\ket0+e^{i\phi}\beta\ket1\).
One has to:\newline (1) Produce the entangled state
\(\alpha\ket0\ket0+\beta\ket1\ket1\)\newline (2) Measure the first
qubit in the basis \(\eta_{\pm}=\left(\ket0\pm
e^{i\phi}\ket1\right)/\sqrt2\).

The result of this measurement is \(\eta_{\pm}\otimes(\alpha\ket0\pm
e^{i\phi}\ket1)=\eta_{\pm}\otimes\psi_{\pm}\) according to whether
the measurement projected the first qubit to \(\eta_+\) or
\(\eta_-\). Now the second tensor factor in this state is the
desired state \(U\psi\) in case of the \(\eta_+\) result, but in the
case of  \(\eta_-\) it is not. Thus we cannot really {\em
implement\/} unitary maps  by measurements and classical
communication, but that is not the point. We can simulate their
action if by simulation we mean that we also include the result of
the final measurement. In quantum computation one makes a final
query of the system to get the result. So what we really want is not
to be able to produce \(U\psi\) but to reproduce the transition
probabilities \(|\melement \omega{U}\psi|^2\) for any state
\(\omega\). Assume that \(\omega\) is \((a\ket0+e^{i\tau}b\ket1)\).
The final step is now:\newline (3) Mesure the resulting state in a
basis that reproduces the desired transition probability.

For this to be possible, one has to have classical communication
from the first measurement.
 If the
first measurement projected the first qubit onto \(\eta_+\) we just
measure the second qubit with \(\ket\omega\bra\omega\) to obtain the
transition probability, but in the case \(\eta_-\) we would get
wrong numbers. However, by classical communication we know which of
the  two cases resulted in the first measurement. If the case was
\(\eta_-\) then we can measure the projector onto the state
\(\tilde\omega=(a\ket0-e^{i\tau}b\ket1)/\sqrt2.\) An easy
calculation shows that \(\bracket\omega{\psi_+}=\melement
\omega{U}\psi=\bracket{\tilde\omega}{\psi_-}\) hence by measurement
and classical communication alone we can reproduce the matrix
elements of unitary operators.  A scheme similar to this, though with other intent, was also introduced by Pati \cite{pati:PRA63.014302}. Of course the above was just a
sketch, but the result is general as shown by the one-way
computation literature \cite{brow-etal}.

Now to be sure, one has to produce the  entangled state in step (1).
This can be done, given \(\psi\), with measurements and classical
communication if one has as an initial resource a certain entangled
states of several qubits. See again \cite{brow-etal}. This initial
entangled state as a resource for all subsequent computation steps
is an earmark of this type of processing. This resource gets
depleted by decoherence resulting from measurements as the
computation progresses, so the process in not reversible, hence the
moniker ``one-way".

Now since all that any experiment does is establishes transition
probabilities and one has no access to the state except through
measurements, one cannot know, still considering our simple case, whether what one is really measuring is \(|\bracket\omega{\psi_+}|^2\) or \(|\bracket{\tilde\omega}{\psi_-}|^2\) as both are equal to \(|\melement \omega{U}\psi|^2\) which is what we are really after. One can thus ask if what is really going on in quantum experiments is a simulation of unitary transformations through some background one-way processes as exemplified by one-way quantum computation. To make this idea clearer we quote the following popular dictum of animal  psychology: {\sl under rigorously controlled laboratory conditions, a rat does what it damn well pleases\/}. Taken over to experimental physics this becomes: {\sl under rigorously controlled laboratory conditions, a measuring instrument does what it damn well pleases.\/}

Suppose that all that exists are measurements and classical
communications. As an experimentalist you prepare a state and
subject it to operations and in the end set up your measuring
apparatus to measure the outcome in a basis \(\phi_1,\dots,\phi_n\).
Now the measuring instrument has received classical communication
from past causally related events, so it does what it damn well
pleases and, taking into account your settings and the communication
received, actually performs a measurement in an alternate basis
\(\tilde\phi_1,\dots,\tilde\phi_n\). This basis is chosen in such a
way as to reproduce exactly the transition probabilities of a
unitary theory. Unitary time evolution is thus subsumed into
measurements and classical communication.

 Does this ontology hold water? Well, to the extend that
one-way computation is feasible, it does. This is already a good
indication of its probable consistency. Some further considerations
are still in order.

First of all, there's the initial entangled state. One has to
suppose that this is the initial state of the universe. We won't go
into how this state has come to be, just as any other ``singular
beginning" for the universe, this is merely a hypothesis without
further explanation. This initial resource gets used up as time
progresses and the fundamental degrees of freedom of the universe
get more and more disentangled. An intrinsic and natural arrow of
time is part of this view.

One may wonder ``what about EPR correlations"? If what we have are
classical events, how can these correlations exist? Don't they now
have to satisfy Bell's inequalities? Well, the state has to be
created by some measurement event which may not {\em really\/}
correspond to the EPR state the experimenter intends, by the
laboratory rat principle. The information as to which state was
created is now broadcast by this event. What we must not assume is
that this event is ``local" in the customary sense, again by the rat
principle. In a sense the ``events" are the ingredients of an
emergent space-time and for this ontology to work one has to assume
that locality notions are constructed from the events, and the
experimenter has no control, by the rat principle, of what is
``local" and what is not. So, the created state now suffers
ostensibly a unitary time evolution which brings it to the measuring
apparatus at both arms of the EPR experiment. This is done by the
one-way quantum computation scheme through measurement events. The
measuring apparatus, having received all the classical information
released by the creation event, and the events of the ostensibly
unitary evolution, now react, being the self-respecting laboratory
rat that it is, in such a way as to reproduce exactly the EPR
quantum correlations. If this is a hidden-variable theory, and it is
not clear that it should be considered as such, it seems it must be
non-local.

\section{Acknowledgements}
This research was partially supported by the Conselho Nacional de Desenvolvimento Cient\'{\i}fico e Tecnol\'ogico (CNPq), and the Funda\c{c}\~ao de Amparo \`a Pesquisa do Estado do Rio de Janeiro  (FAPERJ).

\end{document}